\documentclass{emulateapj}

\usepackage{amsmath}
\usepackage{epsfig}
\usepackage{amssymb}
\usepackage{graphicx}
\usepackage{color}
\usepackage{natbib}

\def\gtaprx {\lower .1ex\hbox{\rlap{\raise .6ex\hbox{\hskip .3ex
	{\ifmmode{\scriptscriptstyle >}\else
		{$\scriptscriptstyle >$}\fi}}}
	\kern -.4ex{\ifmmode{\scriptscriptstyle \sim}\else
		{$\scriptscriptstyle\sim$}\fi}}}
\def\ltaprx {\lower .1ex\hbox{\rlap{\raise .6ex\hbox{\hskip .3ex
	{\ifmmode{\scriptscriptstyle <}\else
		{$\scriptscriptstyle <$}\fi}}}
	\kern -.4ex{\ifmmode{\scriptscriptstyle \sim}\else
		{$\scriptscriptstyle\sim$}\fi}}}

\newcommand{\cutt}[1]{\textcolor{blue}{}}

\newcommand{\Ms}{{\ensuremath{M_{\odot} }}}
\newcommand{\Zs}{\ensuremath{Z_\odot}}

\begin{document}

\title{Radiation Hydrodynamical Simulations of the First Quasars}

\author{Joseph Smidt\altaffilmark{1}, Daniel J. Whalen\altaffilmark{2}, Jarrett L. 
Johnson\altaffilmark{1}, Marco Surace\altaffilmark{2} and Hui Li\altaffilmark{1}}

\altaffiltext{1}{Los Alamos National Laboratory, Los Alamos, NM 87545}

\altaffiltext{2}{Institute of Cosmology and Gravitation, Portsmouth University, Dennis 
Sciama Building, Portsmouth PO1 3FX, UK}

\begin{abstract}

Supermassive black holes (SMBHs) are the central engines of luminous quasars and are found in most massive galaxies today. But the recent discoveries of ULAS J1120+0641, a $2 \times 10^9$ \Ms\ BH at $z =$ 7.1, and ULAS J1342+0928, a $8.0 \times 10^{8}$ \Ms\ BH at $z =$ 7.5, now push the era of quasar formation up to just 690 Myr after the Big Bang.  Here we report new cosmological simulations of SMBHs with X-rays fully coupled to primordial chemistry and hydrodynamics that show that J1120 and J1342 can form from direct collapse black holes (DCBHs) if their growth is fed by cold, dense accretion streams, like those thought to fuel rapid star formation in some galaxies at later epochs.  Our models reproduce all of the observed properties of J1120:  its mass, luminosity, and H II region as well as star formation rates and metallicities in its host galaxy.  They also reproduce the dynamical mass of the innermost 1.5 kpc of its emission region recently measured by ALMA and J-band magnitudes that are in good agreement with those found by the VISTA Hemisphere Survey.  
 
\end{abstract}


\maketitle

\section{Introduction}

Nearly 160 quasars have now been discovered at $z >$ 6, including ULAS J1120+0641, a $2 \times 10^9$ \Ms\ BH at $z =$ 7.1, and ULAS J1342+0928, an $8.0 \times 10^{8}$ \Ms\ BH at $z =$ 7.5 \citep{mort11,ban18}.  Their origins and how they grew to such large masses by such early times are not yet clear, but they might have formed from either Population III star black holes (Pop III BHs) at $z \sim$ 25 or direct-collapse black holes (DCBHs) at $z \lesssim$ 20 \citep{km05,jet13}.  Pop III BHs are born in low densities that preclude rapid initial growth, so they would later require extended periods of hyper-Eddington accretion to reach 10$^9$ \Ms\ by $z \sim$ 7 to overcome the low duty cycles imposed by radiation from the BH \citep[e.g.,][]{wan04,milos09a,milos09b,pm11,pm12,pm13,mhd14,vsd15}.  However, even if such modes of accretion could occur it is not clear that the BH would encounter enough gas to become an SMBH by $z \sim$ 7 \citep{awa09,srd18}. Many low-mass Pop III BH seeds would also be ejected from their birth halos, and thus their fuel supplies, by asymmetries in their core-collapse engines \citep{wf12}.  

On the other hand, rapid baryon collapse in atomically-cooled halos \citep{ln06,wta08,rh09b,sbh10,latif13a} and the formation of supermassive stars that collapse to DCBHs \citep{hos13,um16,tyr17,hle18,hle17} require unusually strong sources of Lyman-Werner (LW) UV photons \citep{agarw12,dfm14,yue14,jet14,regan17a,chon17a,rd18} or supersonic baryon streaming motions that may or may not occur in any given halo (\citealt{greif11,stacy11a,lns14,hir17,srg17} -- see also \citealt{io12,ivk15}).  But DCBHs can grow at much higher rates than Pop III BHs because they are born with much larger masses in much higher densities in host galaxies that can retain their fuel supply, even when heated by X-rays.  A DCBH may now have been discovered in the Ly-$\alpha$ emitter CR7 at $z =$ 6.6 \citep{cr7,til15b,smidt16a}.
 
To reach 10$^9$ \Ms\ by $z \sim$ 7, SMBH seeds must also be fueled for long times, either by a succession of mergers with other gas-rich halos \citep[e.g.,][]{li07} or by unusually strong cold accretion flows, like those thought to drive rapid star formation in some galaxies at later epochs \citep[e.g.,][]{dek06,dek09,bour11}.  There may be up to a few dozen regions per Gpc$^{-3}$ with cold flows capable of forming quasars by $z \gtrsim$ 7 (\citealt{dm12,yu14}; see also \citealt{costa14,hir14}).  

After birth, X-rays from the nascent BH regulate its growth by heating and ionizing flows in its environment and photoevaporating cold streams outside the halo.  Winds, ionizing UV and supernovae (SNe) from stars later perturb flows onto the BH, either promoting or suppressing its growth \citep[see, e.g.,][]{dub15,hab17}.  Stars also govern their own rates of formation because their ionizing UV can trigger or quench star formation in nearby clouds  \citep{oet05,wet08b,wet10}.  Metals from SNe also cool gas and create new stars \citep{ky05,wet08a,ritt12,ss13,brit15,rit16,slud16,chen17a,chen17b}.  

Radiation from the BH can also promote or suppress star formation in its vicinity  \citep{mba03,aycin14}.  X-rays produce energetic photoelectrons that cause secondary ionizations that enhance free electron fractions in primordial gas.  Free electrons in turn catalyze the formation of H$_2$ via the H$^{-}$ channel, cooling and collapsing gas and creating more stars \citep[e.g.,][]{ga08}.  On the other hand, radiation from the BH can also evaporate star-forming clouds and global LW backgrounds can destroy H$_2$, quenching star formation.  

To model the formation of primordial quasars, simulations must resolve all these processes deep in the host galaxy of the BH and in cold streams in the intergalactic medium (IGM).  We have now bridged these scales with new simulations that include X-rays from the BH and winds, ionizing UV and SNe from star-forming regions, and have evolved a quasar in cold flows from birth at $z = 19$ down to $z = 6$.  We describe these models in Section 2 and examine the evolution of the BH and its host galaxy in Section 3, where we also compare its properties with those observed for J1120.  We discuss our results and conclude in Section 4.

\section{Numerical Method}

We use the Enzo adaptive mesh refinement (AMR) cosmology code \citep{enzo} with the MORAY radiation transport package \citep{moray} to model the quasar in this study.  X-ray and ionizing UV transport in MORAY includes radiation pressure on gas due to photoionizations and is self-consistently coupled to hydrodynamics and nine-species nonequilibrium primordial gas chemistry in Enzo.  Secondary ionizations due to energetic photoelectrons and Compton heating by X-rays are taken into account in the chemistry and energy equations along with the usual primordial gas cooling processes:  collisional excitational and ionizational cooling by H and He, recombinational cooling, bremsstrahlung cooling, H$_2$ cooling, and inverse Compton cooling by the cosmic microwave background.  

The BH is represented by a modified star particle whose X-ray luminosity, $L_{\mathrm{r}}$, is $\epsilon_{\mathrm{r}} \dot{m}_{\mathrm{BH}} c^2$, where $\epsilon_{\mathrm{r}}$ is the mean radiative efficiency, taken to be 0.1, and $\dot{m}_{\mathrm{BH}}$ is the accretion rate.   This luminosity is all in the form of 1 keV photons to maximize the heating of gas because at higher energies they have smaller ionization cross sections and at lower energies they deposit less energy per ionization \citep{1kev,hum16}.  This energy is also consistent with new observations showing 90\% of the X-ray flux of J1120 to be at 0.5 - 2 keV \citep{nan17}.  Our simulations do not resolve the accretion disk of the BH, which is $\sim$ 1 pc in diameter, so we use an alpha disk model to compute accretion rates to approximate angular momentum transport out of the disk on subgrid scales \citep{alphad}.  We also approximate a disk wind by depositing 10$^{-4}$ $L_r$ as thermal energy above and below the midplane of the BH perpendicular to its angular momentum vector \citep{ciotti09}.

\begin{figure}
\begin{center}
\includegraphics[width=\columnwidth]{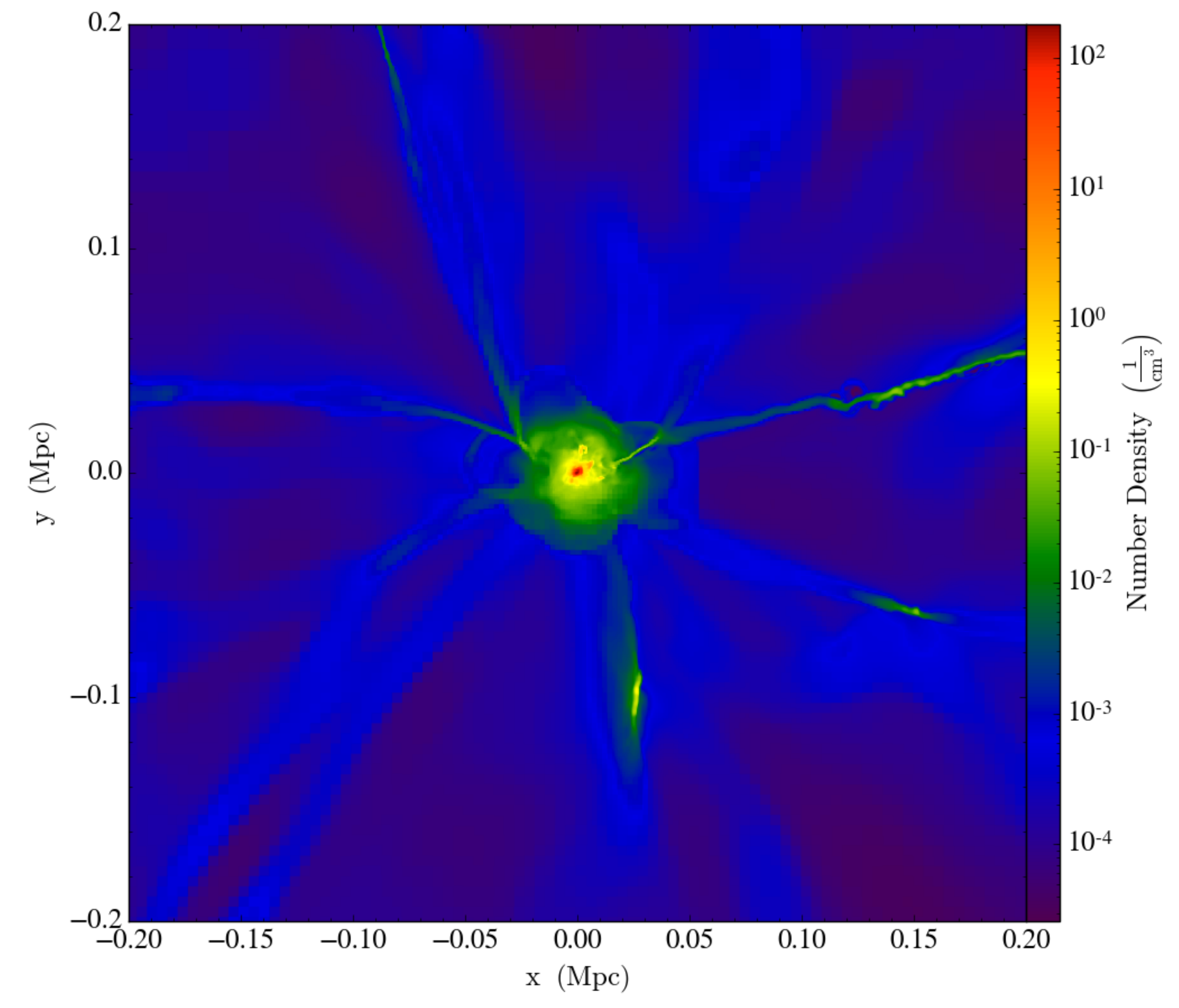} 
\caption{Density slice of the halo at $z = 7.1$.  Cold accretion streams intersecting the host galaxy of the quasar are clearly visible. Distances are in comoving Mpc. The cold streams have temperatures of $\sim$ 500 K due to H$_2$ cooling.
\label{fig:halo}}
\end{center}
\vspace{-0.1in}
\end{figure}

We use a stochastic prescription for star formation that is based on \citet{co92} (hereafter CO92) but is slightly modified to accommodate a minimum star particle mass to avoid having too many low-mass particles \citep[section 8.2.2 of][]{enzo}.  The algorithm tallies the mass accumulated in a region until it exceeds the minimum star particle mass, at which point it creates a star particle there if the conditions for SF in CO92 have also been met.  This accumulated mass exceeds the minimum mass at random times so particles are formed sporadically.  Furthermore, the particles will have a range of masses because the CO92 criteria might not be satisfied when the accumulated mass in a region exceeds the minimum mass, so it will grow in mass until they are met.  We adopt a minimum star particle mass of 10$^7$ \Ms.

Star particles are assigned a Salpeter initial mass function (IMF) for simplicity and are tagged as sources of ionizing UV photons, which are propagated throughout the simulation box by MORAY.  Each particle emits photons at four energies: 12.6 eV (LW photons), 21.62 eV (the average energy of UV photons from massive, low-metallicity stars), 30 eV (He I ionizing photons), and 60 eV (He II ionizing photons).  The relative numbers of photons apportioned to these bins are determined from the fits for the $Z = 0$, no mass-loss tracks in Table 6 of \citet{s02}. The particles also deposit momentum and metals from stellar winds into the ISM over their lifetimes.  SN feedback is modeled as thermal energy, with 10$^{51}$ erg deposited per explosion assuming one SN per 200 \Ms\ of stars formed.  Cooling by metals from SNe is included with rates from \citet{japp07}, assuming solar yields that are consistent with our chosen IMF.   Our models also have a uniform Lyman-Werner background due to early stars that evolves with redshift.

\begin{figure*}
\begin{center}
\begin{tabular}{cc}
\includegraphics[width=\columnwidth]{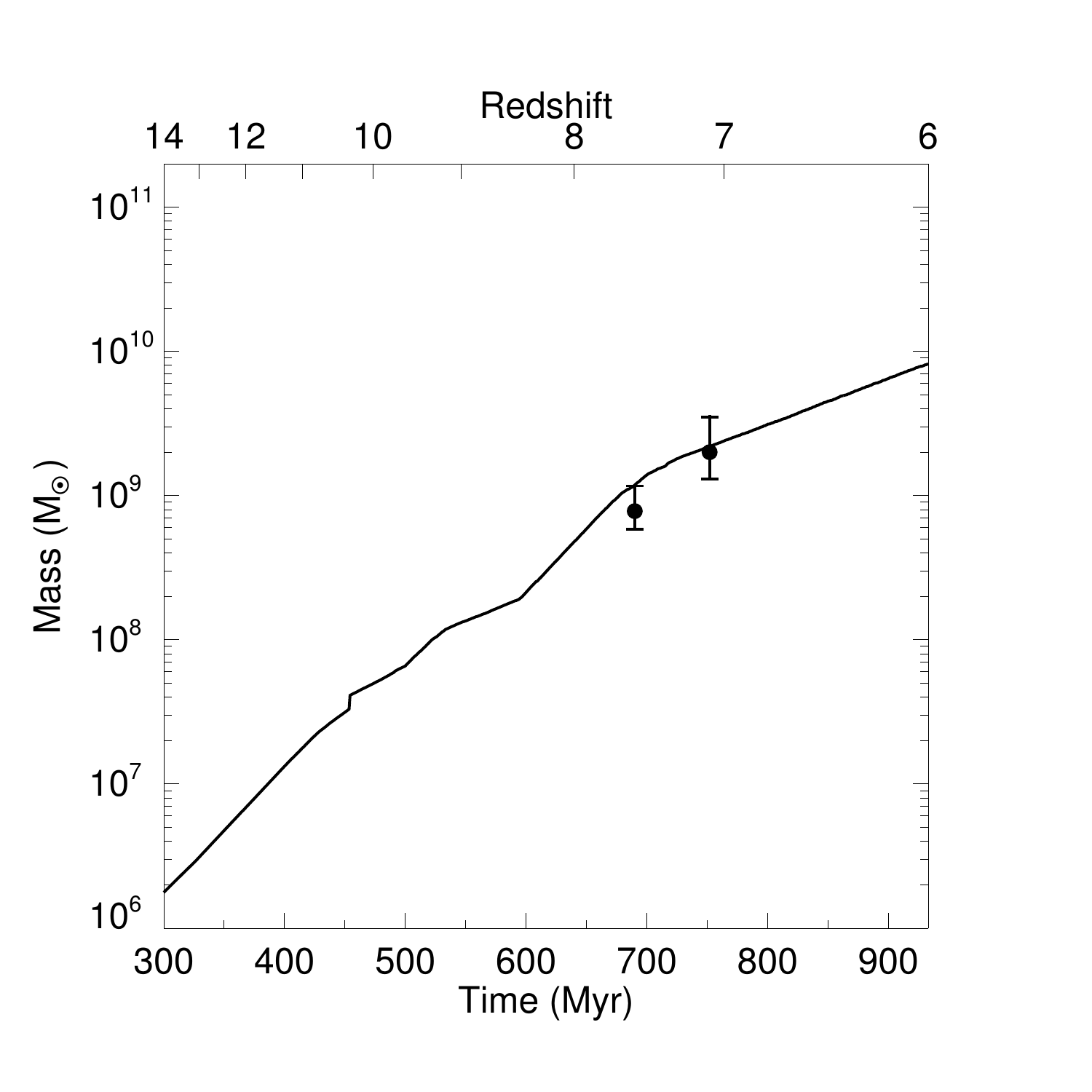} &
\includegraphics[width=\columnwidth]{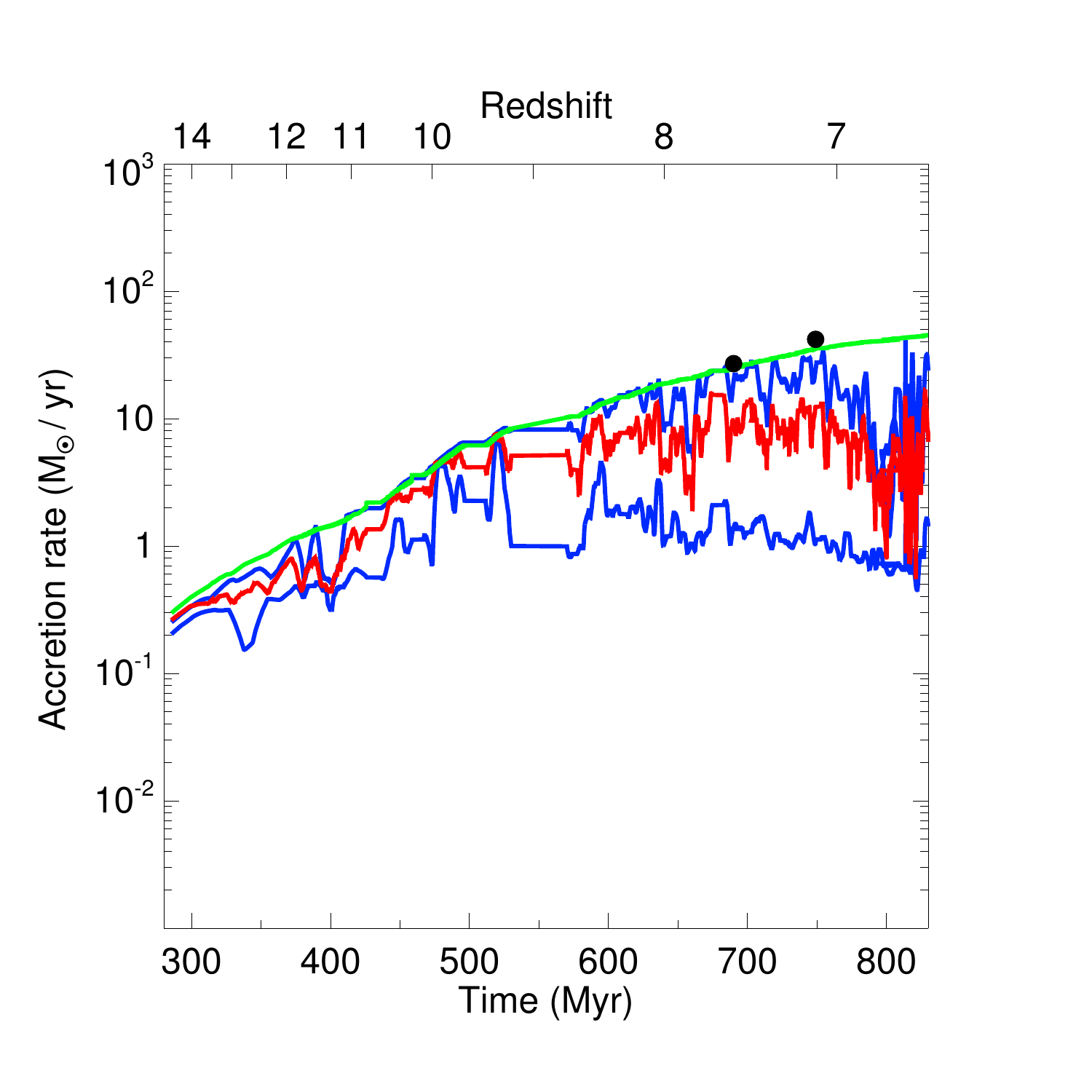} \\
\end{tabular}
\end{center}
\vspace{-0.3in}
\caption{Left panel: BH mass as a function of time and redshift. The solid black circles and their error bars are the masses of J1342 and J1120 from observations at $z =$ 7.5 and 7.1, respectively.  Right panel: average BH accretion rates (red), upper and lower limits on the accretion rate (blue) and the Eddington rate (green). The black circles at $z =$ 7.5 and 7.1 are the accretion rates for the observed bolometric luminosities of J1342 and J1120 respectively, assuming $\epsilon =$ 0.1.}
\label{fig:bhmass}
\end{figure*}

Our simulation box is 100 $h^{-1}$ Mpc on a side, with a 256$^3$ root grid and three nested 25 $h^{-1}$ Mpc grids that are centered on the host halo for an effective resolution of 2048$^3$. These grids yield initial dark matter and baryon mass resolutions of 8.41 $\times$ 10$^6$ $h^{-1}$ \Ms\ and 1.57 $\times$ 10$^6$ $h^{-1}$ \Ms, respectively.  The grid is initialized with gaussian primordial density fluctuations at $z =$ 200 with MUSIC \citep{hahn11} with cosmological parameters from the second-year \textit{Planck} best fit lowP+lensing+BAO+JLA+H$_0$: $\Omega_{\mathrm{M}}=$ 0.308, $\Omega_{\Lambda}=$ 0.691, $\Omega_{\mathrm{b}} = $ 0.0223, $h =$ 0.677, $\sigma_8 = $ 0.816, and $n =$ 0.968 \citep{planck2}.  We use a maximum refinement level $l$ = 10 and refine the grid on baryon overdensities of 3 $\times$ 2$^{-0.2l}$ to obtain a maximum resolution of 35 pc (comoving).  We also refine on a dark matter overdensity of 3 and resolve the local Jeans length with at least 32 zones at all times to avoid artificial fragmentation during collapse.  

The box was chosen to be large enough to enclose the cold flows feeding the quasar on cosmological scales while resolving gas flows, photoionization and star formation deep within its host galaxy. However, given that there are only about a dozen regions per Gpc$^{-3}$ with reservoirs capable of sustaining rapid quasar growth in this manner, no single 100 Mpc box at random would be expected to enclose one.  We therefore tested multiple random seeds at lower resolution until we obtained a halo that exceeded 10$^{12}$ \Ms\ by $z \sim$ 7 and was the product of at most a few major mergers, i.e., with other halos similar in mass to the host halo \citep{tss08}.  The host halo of the BH is shown at $z = 7.1$ in Fig.~\ref{fig:halo} to illustrate the morphology of the cold streams.  We evolve the box from $z =$ 200 to 19.2, when the host halo reaches 3.0 $\times$ 10$^8$ \Ms\ and begins to atomically cool and collapse.  At this redshift we initialize a 10$^5$ \Ms\ BH particle at the center of the halo and turn on X-rays and star formation.    

\section{SMBH / Galaxy Coevolution}

\begin{figure*}
\begin{center}
\begin{tabular}{cc}
\includegraphics[width=\columnwidth]{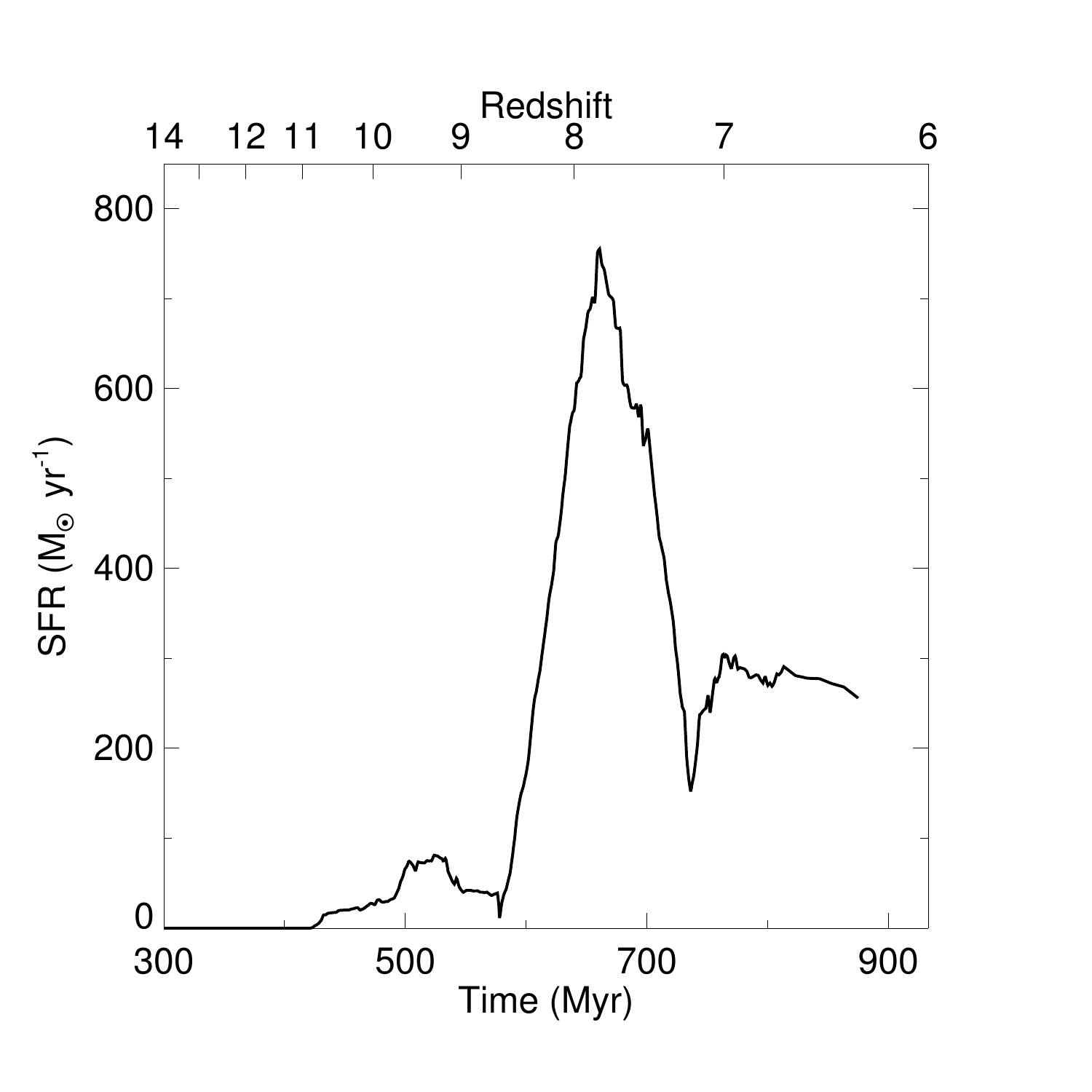} &
\includegraphics[width=\columnwidth]{f3a.pdf} \\
\end{tabular}
\end{center}
\caption{Left panel: Star formation rates in the host galaxy of the BH.  Right panel: BH accretion rates normalized to the Eddington limit.}
\label{fig:sfr}
\end{figure*}

BH masses and accretion rates are shown in Fig.~\ref{fig:bhmass} and star formation rates (SFRs) in the host galaxy and average accretion rates normalized to the Eddington limit are shown in Fig.~\ref{fig:sfr}.  Actual accretion rates fluctuate rapidly between the limits shown in blue, with average values in red.  Prior to the onset of star formation in the host galaxy, X-ray heating by the nascent BH limits average accretion rates to 0.8 $\dot{m}_{\mathrm{Edd}}$, where $\dot{m}_{\mathrm{Edd}}$ is the Eddington accretion limit, but this fraction soon falls to 0.3 by 400 Myr as the BH grows two decades in mass.  X-rays regulate the growth of the BH even at these early times because cold streams carry dense gas deep into its host galaxy, raising its average density to 10 - 100 cm$^{-1}$.  At these densities 1 keV photons have mean free paths of 200 - 300 pc, less than the virial radius of the halo, so they efficiently heat the gas.  

The host halo merges with a $1.2 \times 10^9$ \Ms\ halo at 400 Myr and a $1.0 \times 10^{10}$ \Ms\ halo at 450 Myr (mass ratios of 1:1 and 1:4, respectively).  These collisions trigger the small starburst that begins at 420 Myr and peaks at 540 Myr.  Disruption of the halo by the mergers and SN feedback from the burst perturb flows onto the BH, inducing brief episodes of super-Eddington accretion at 475 Myr and 510 Myr and causing the slight jump in BH mass at 450 Myr ($z =$ 10.7) in Fig.~\ref{fig:bhmass}. The accretion ratio again begins to fall after this first starburst subsides.

After the two major mergers, the halo grows primarily by accretion from $5 \times 10^9$ \Ms\ at $z \sim 11$ to $1.2 \times 10^{12}$ \Ms\ by $z = 7.1$.  However, we note that while the halo mass does not exhibit any later jumps indicative of major mergers, smaller halos merge with the host halo over its entire evolution and are also carried into it by the cold streams themselves.  After the first starburst, metals injected into the ISM by SNe trigger gas cooling and collapse that lead to a second, much larger starburst at 580 Myr that peaks at 770 \Ms\ yr$^{-1}$ ($z =$ 7.8).  The absence of jumps in halo mass over this period indicates that the burst is triggered by metal cooling and primarily fueled by accretion.

The much stronger SN feedback from this burst and larger X-ray fluxes from the now more massive BH cause average accretion efficiencies to fall after 500 Myr, eventually to 0.2 $\dot{m}_{\mathrm{Edd}}$ by 900 Myr.  Even so, the upper limit to the fluctuations in these rates is still at nearly the Eddington limit at $z =$ 7.1, consistent with observations of J1120.  At no time do we artificially cap accretion rates onto the BH; except for brief episodes, X-rays and SN feedback limit them to below the Eddington rate throughout its life.  The second, much larger burst in star formation is quenched by its own UV and SN feedback, falling to $\sim$ 150 \Ms\ yr$^{-1}$ by $z \sim$ 7.3.  The SFR then recovers but levels off, never again exceeding $\sim$ 300 \Ms\ yr$^{-1}$ because of rising X-ray fluxes from the BH, even though cold streams continue to flow into the halo as shown in Fig.~\ref{fig:halo}.  

\begin{figure*}
\begin{center}
\begin{tabular}{cc}
\epsfig{file=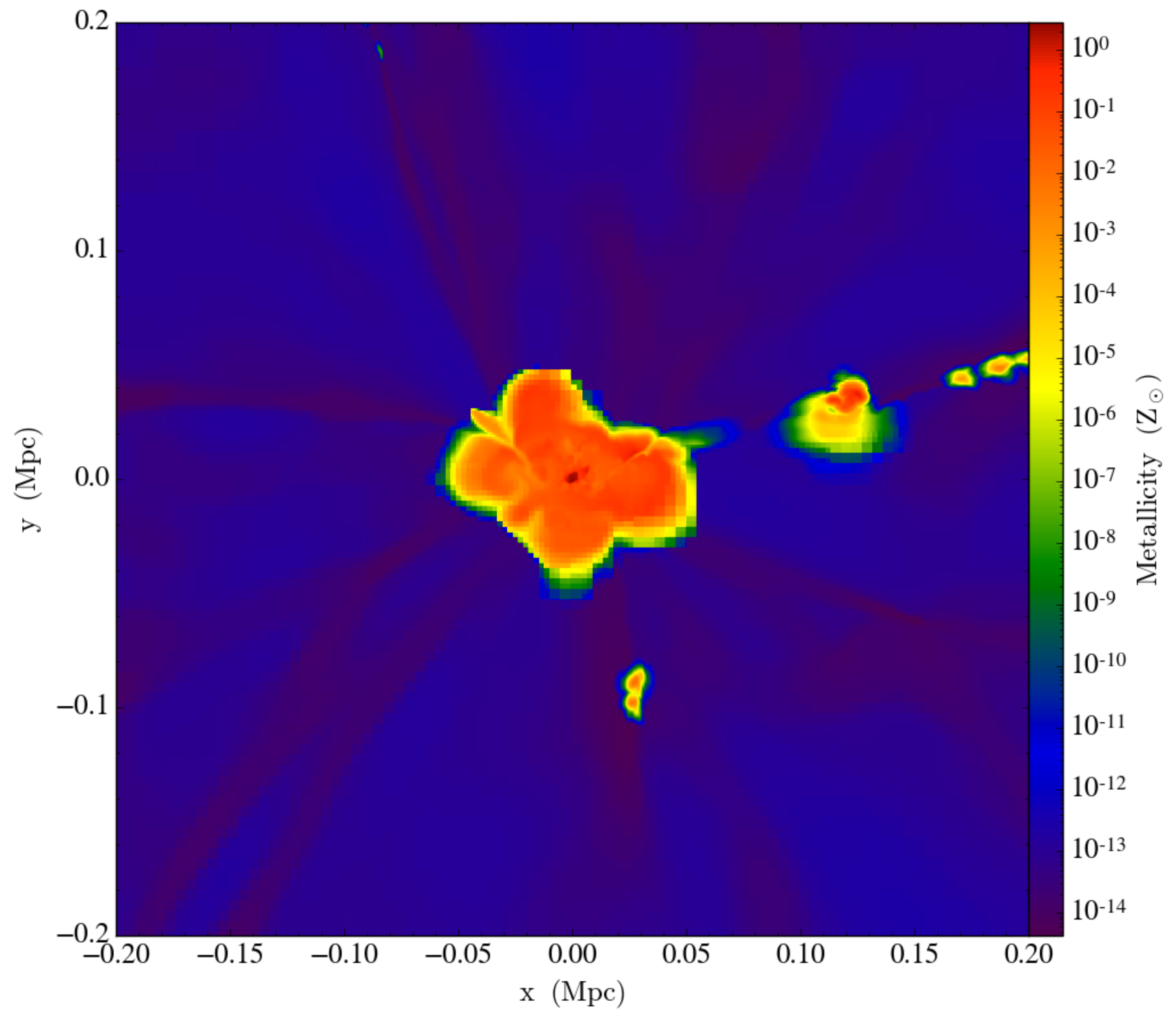,width=0.45\linewidth,clip=} & 
\epsfig{file=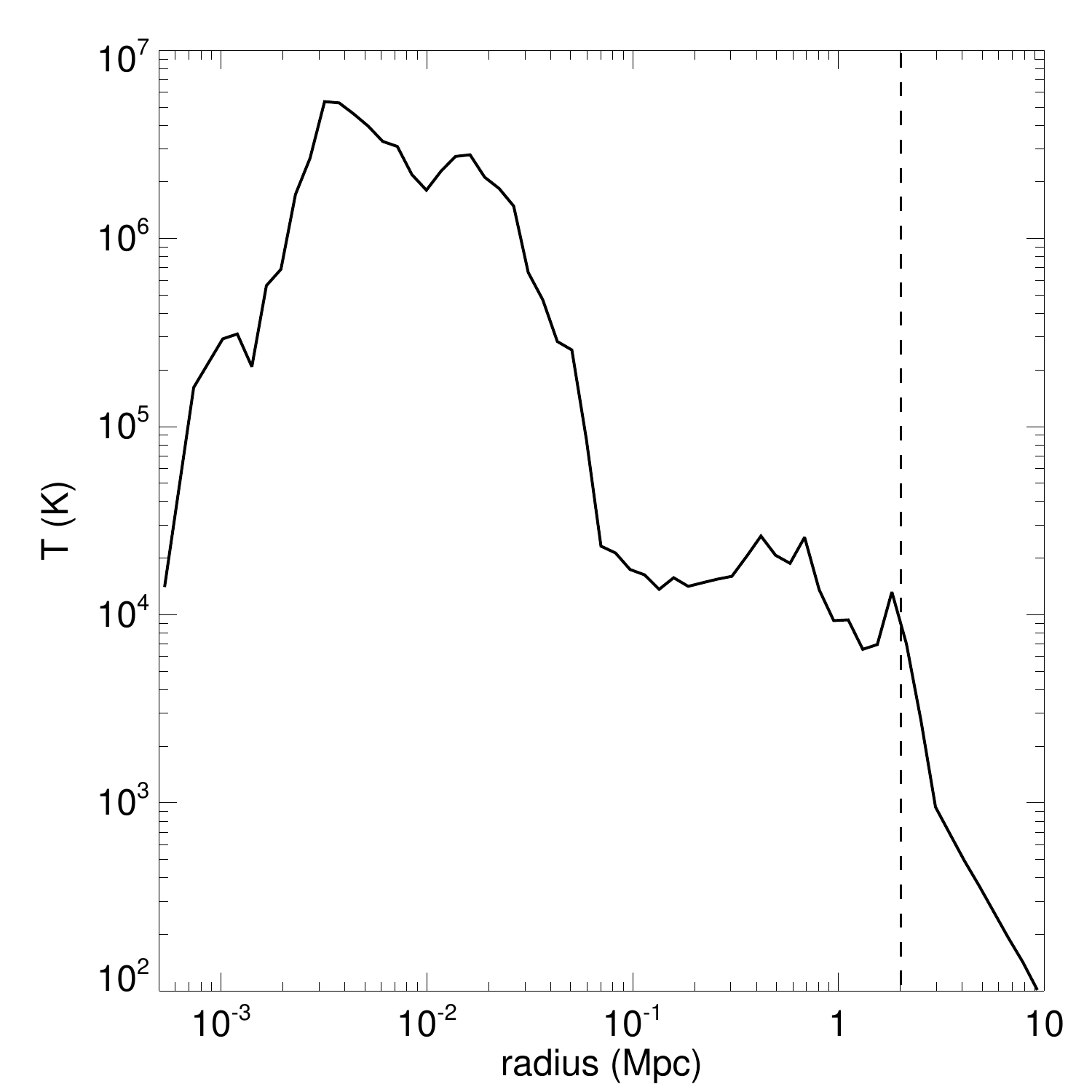,width=0.40\linewidth,clip=} \\
\end{tabular}
\end{center}
\caption{Left: metallicity slice through the center of the host galaxy of the quasar at $z =$ 7.1. Distance scales are in comoving Mpc. Right:  Spherically averaged temperature profile of the H II region of the quasar at $z =7.1$. The vertical line marks the approximate boundary of gas at $>$ 10$^4$ K at $\sim$ 2 Mpc, the observed radius of the ionized near zone of ULAS J1120+0641.}
\label{fig:metals}
\end{figure*}

At $z = 7.1$ the BH mass is $2.15 \times 10^9$ \Ms, within the error bars in mass for J1120.  The SFR in the host galaxy is 245 \Ms\ yr$^{-1}$, consistent with observations \citep[105 - 340 \Ms\ yr$^{-1}$;][]{vet17}.  The dynamical mass of the central 1.5 kpc of the host galaxy is $3.95 \times 10^{10}$ \Ms, in good agreement with recent measurements of (4.3 $\pm$ 0.9) $\times$ 10$^{10}$ \Ms\ by the Atacama Large Millimeter Array \citep[ALMA;][]{vet17}.  As shown in Fig.~\ref{fig:metals}, metallicities are mildly supersolar ($\sim$ 2 \Zs) at the center of the host galaxy and approximately solar throughout the rest of its interior, as found in observations \citep[Fig.~2 in][]{dunlop13}.  Metals are also visible in nearby halos and in some of the accretion filaments threading the host galaxy.  The spherically-averaged temperature profile of the H II region of the quasar at $z = 7.1$ is shown in the right panel of Fig.~\ref{fig:metals}. Both X-rays from the BH and SNe heat gas in the host galaxy to a few 10$^6$ K and the X-rays heat the surrounding IGM to temperatures of 10$^4$ - 10$^5$ K.  Temperatures of 10$^4$ K extend out to $\sim$ 2 Mpc from the quasar, which is the observed radius of the ionized near zone of J1120.  Using the CLOUDY code \citep{cloudy} to compute the NIR luminosity of the quasar, we obtain a J band AB mag $=$ 20.6 at $z =$ 7.1, in good agreement with 20.2 from the VISTA High Latitude survey \citep[VHS;][]{mort11,reed17}.

\begin{figure*}
\begin{center}
\begin{tabular}{cc}
\epsfig{file=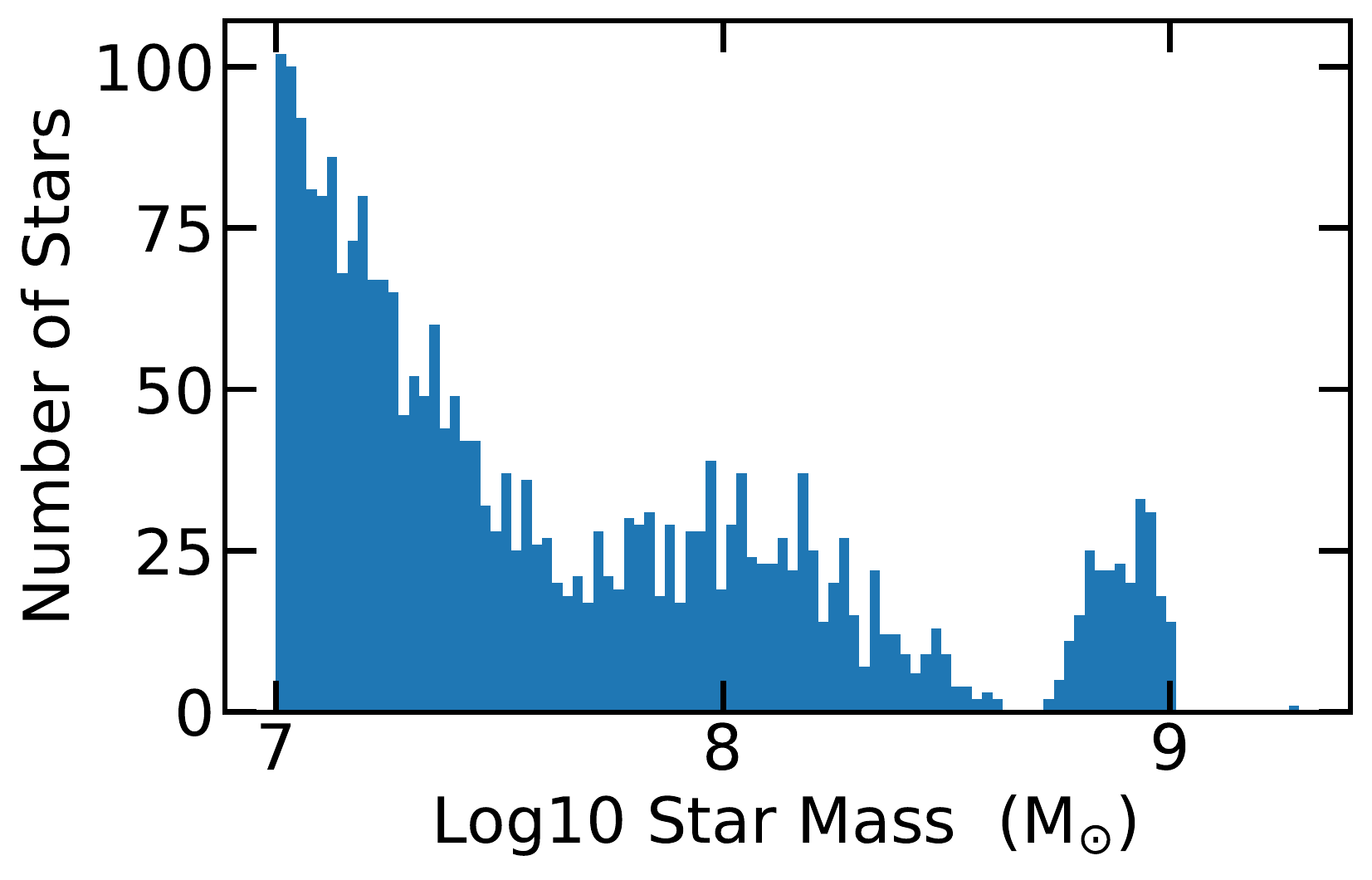,width=0.475\linewidth,clip=} & 
\epsfig{file=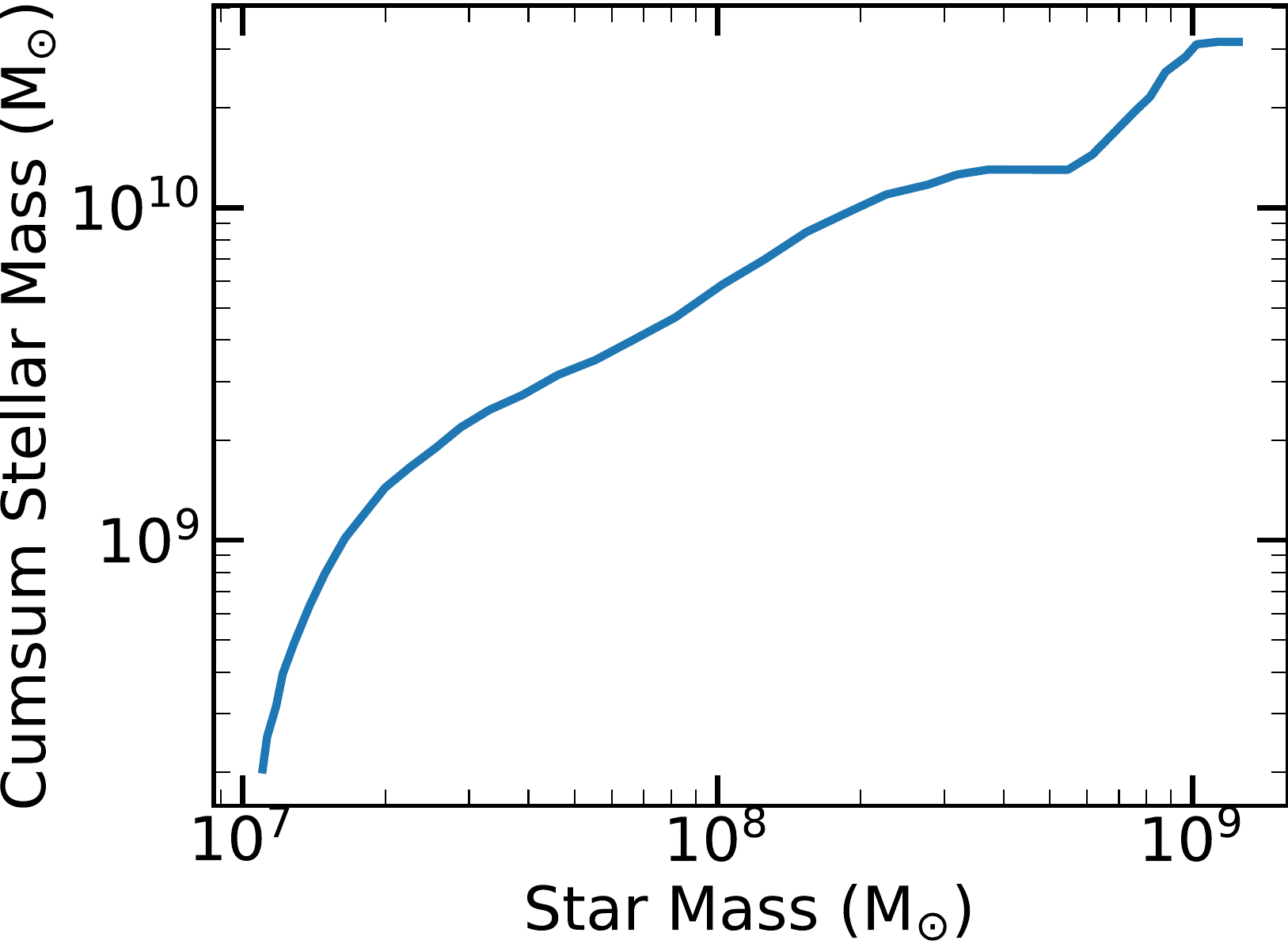,width=0.475\linewidth,clip=} \\
\end{tabular}
\end{center}
\caption{Stellar mass of the host galaxy of the quasar at $z =$ 7.1.  Left panel: number of star particles binned by log mass.  Right panel: cumulative stellar mass as a function of star particle mass.}
\label{fig:SPs}
\end{figure*}

We show a census of star particles for the host galaxy of the quasar at $z =$ 7.1 in the left panel of Figure~\ref{fig:SPs}.  The masses of the particles range from the minimum of 10$^7$ \Ms\ up to a few 10$^9$ \Ms.  While there are a large number of particles with masses of 10$^{7 - 7.5}$ \Ms\ most of the stellar mass of the galaxy is in the 10$^{8 - 9}$ \Ms\ particles, as can be seen in the right panel of Figure~\ref{fig:SPs}.  The total mass of the stars is 3.1 $\times$ 10$^{10}$ \Ms, or 1.6\% of the mass of the halo.  This ratio is consistent with Figure 9 of \citet{bh14}, who predict that the stellar mass of a 10$^{12}$ \Ms\ halo at $z =$ 7 should be 1 - 2\% of the mass of the halo.  There is a total of 2641 star particles in the host galaxy at this redshift.

\begin{figure*}
\begin{center}
\begin{tabular}{cc}
\epsfig{file=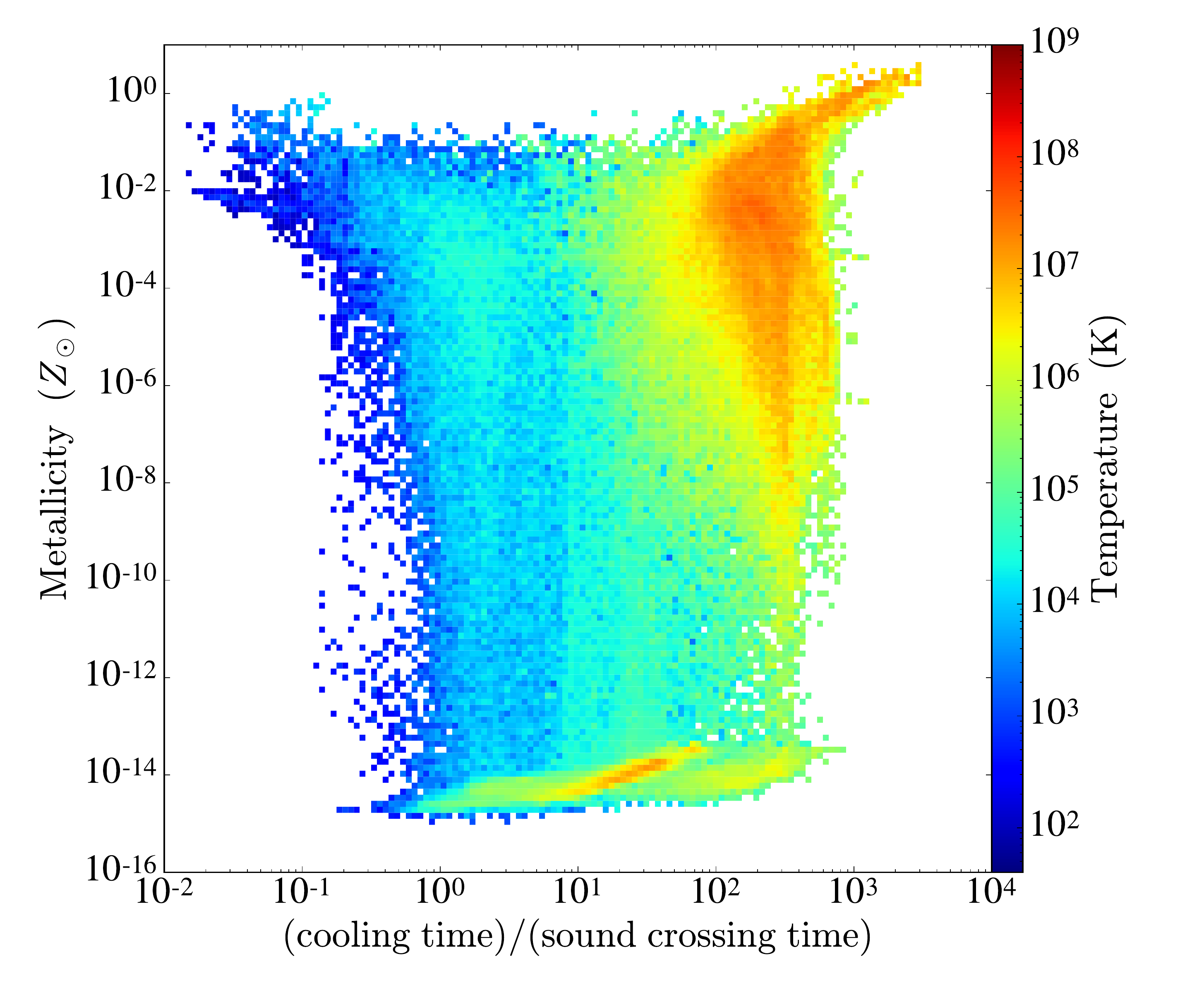,width=0.475\linewidth,clip=} & 
\epsfig{file=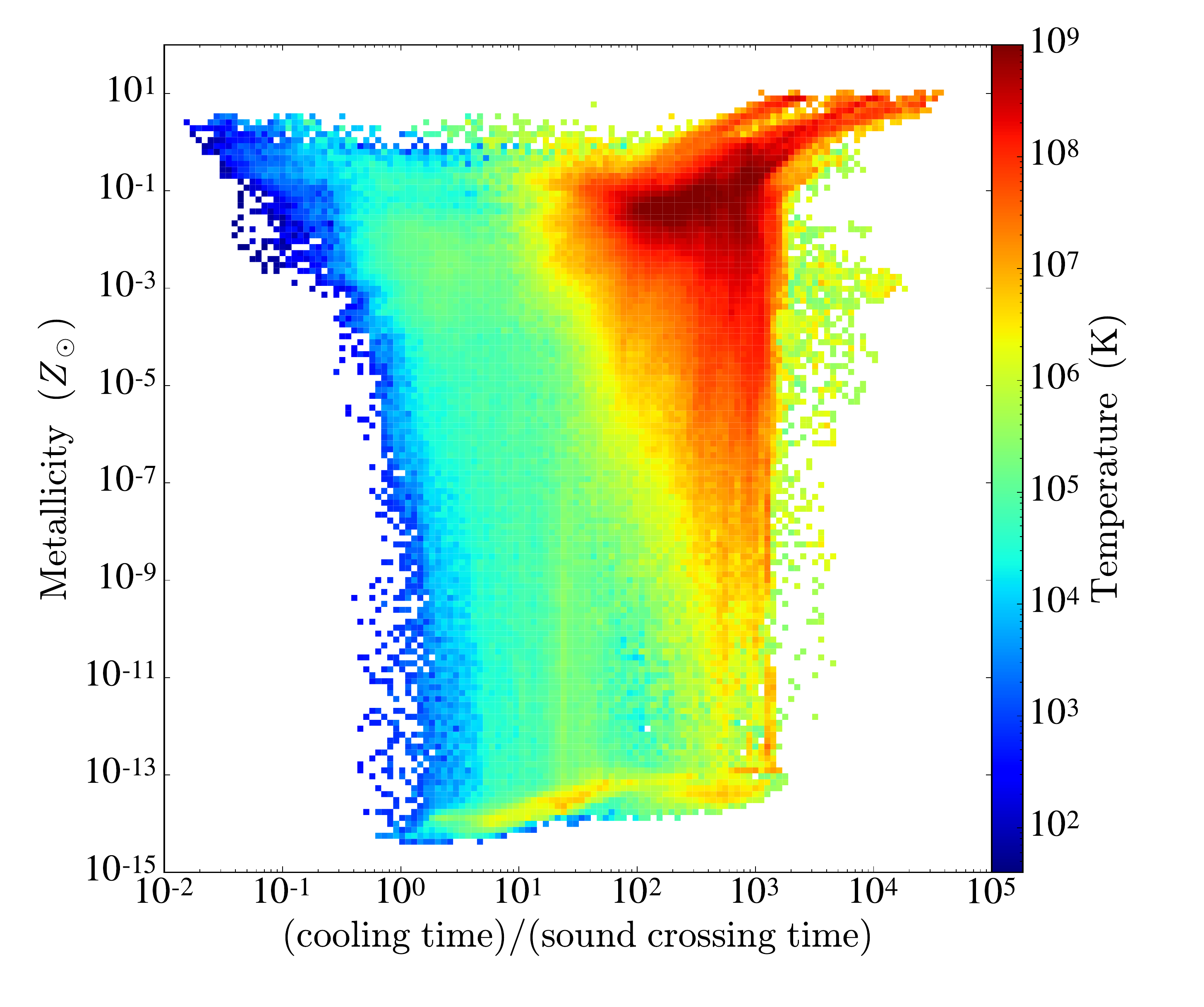,width=0.475\linewidth,clip=} \\
\end{tabular}
\end{center}
\vspace{-0.3in}
\caption{Ratio of the cooling time to the sound crossing time in the host galaxy of the SMBH.  In regions of high metallicity and temperatures (10$^7$ K), in which thermal energy due to SNe is injected, these ratios are always much larger than 1.  Left panel: 450 Myr, the beginning of the first starburst.  Right panel: 750 Myr, the end of the second starburst.}
\label{fig:dvs12}
\end{figure*}

SN feedback can be unphysically quenched in numerical simulations if the energy of the explosion is deposited as heat in star forming regions with high densities. Rapid cooling can dissipate this energy before it can launch flows that disrupt star formation or the growth of the BH.  But as we show in Fig.~\ref{fig:dvs12}, this never happens in our models: the ratio of the cooling time to the sound crossing time \citep[equation 19 in][]{dvs12} in regions of high metallicities and temperatures ($> 10^7$ K) in the host galaxy is always greater than 1 and usually above 10.  These ratios are due to the injection of momentum into the ISM by stellar winds and ionization heating by UV from stars, both of which clear gas away from star particles prior to explosion. Thermal energy from SNe is therefore deposited in lower densities that cannot cool quickly.  X-rays from the BH also permeate the host galaxy in our simulations, heating gas and expanding it to lower densities.

\section{Discussion and Conclusion}

\begin{figure}
\begin{center}
\includegraphics[width=\columnwidth]{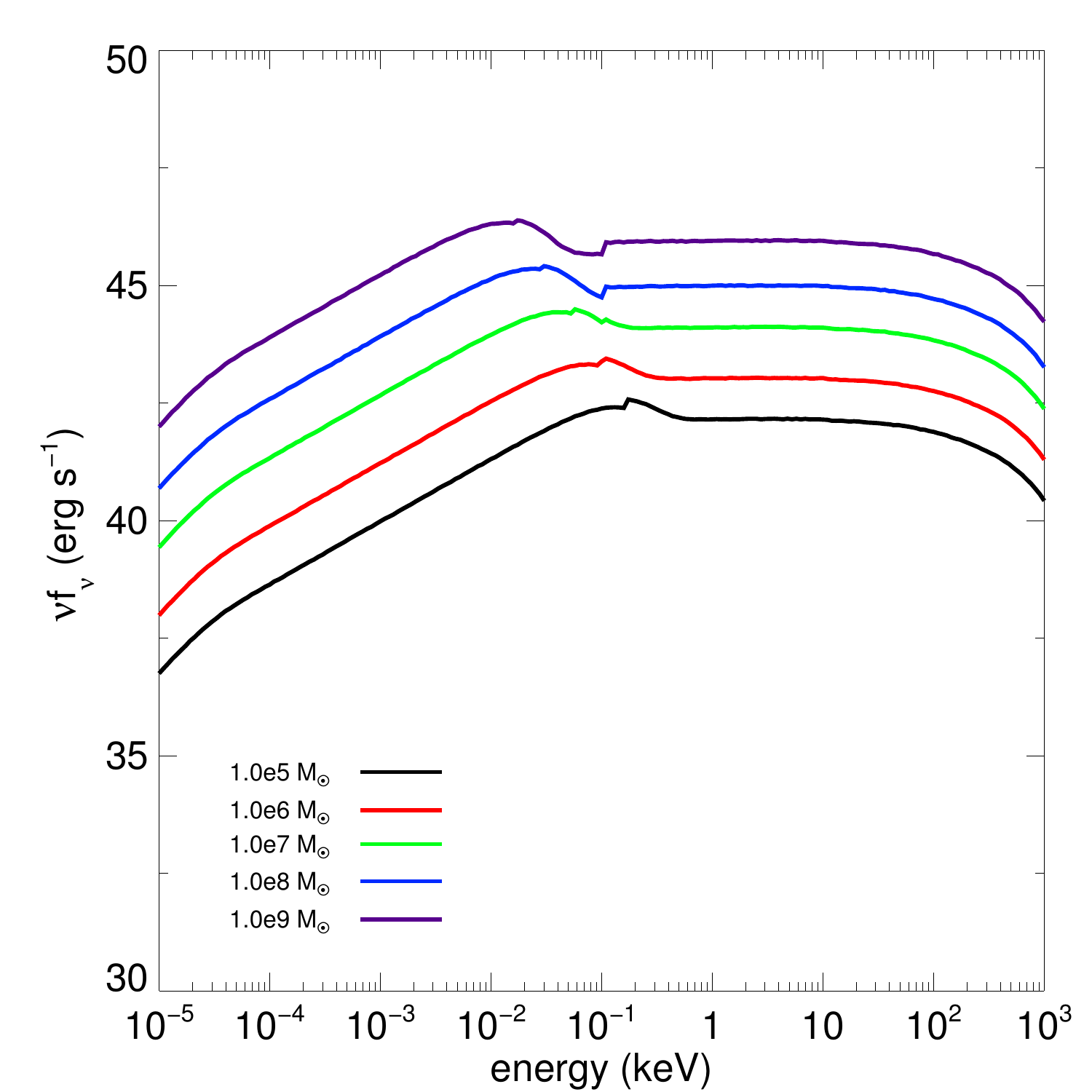} 
\caption{Theoretical BH spectra from 10$^5$ - 10$^9$ \Ms\ from \citet{pac15}.
\label{fig:BHspec}}
\end{center}
\vspace{-0.1in}
\end{figure}

We adopted 1 keV photons for the BH spectrum in our models in part because it is the band in which most X-ray energy is found in high-$z$ quasars \citep{nan17}.  This choice produced good fits to all observed properties of J1120 at $z \sim$ 7 in our model.  However, actual quasar spectra are expected to evolve from X-rays down to extreme UV over the large range in BH mass in our study.  We show theoretical spectra for 10$^5$ - 10$^9$ \Ms\ BHs from \citet{pac15} in Fig.~\ref{fig:BHspec}.  The peak in spectral emission moves steadily towards longer wavelengths as the BH grows in mass, but the spectral energy distribution itself remains flat from 0.5 - 10 keV.  These models neglect Compton upscattering by hot gas in the disk, which would redistribute photons at the spectral peak into the keV range \citep{done12}.  The average photon energy of these spectra ranges from 2.0 keV at 10$^5$ \Ms\ to 0.3 keV at 10$^9$ \Ms, so our choice of 1 keV is an appropriate average for this interval in BH mass.

That said, lower-energy photons in actual SMBH spectra may have a measurable effect on the evolution of the quasar because they have larger cross sections to ionization in H, He and some metal ions.  But this effect appears to be modest at lower BH masses.  \citet{smidt16a} evolved a DCBH over two decades in mass with both monoenergetic X-ray and multigroup spectra that included ionizing UV and found little change in its evolution. Radiation pressure from IR on dust in the host galaxy of the BH may also drive strong but short-lived  outflows \citep{bieri17,costa18a,costa18b}, although such outflows have only been observed in a fraction of quasars at $z > $ 6.  The softer components of SMBH spectra may also quench their images at 21 cm in emission because they ionize rather than heat the IGM.  The effects of evolving BH spectra on the growth of the first quasars will be examined in future simulations.

While our models trace metal production due to star formation, they do not include dust formation and cooling or opacity due to dust and metals in the radiation transport.  Dust cooling would promote star formation in environments in which radiation does not destroy it first, and both metals and dust would enhance X-ray heating in the galaxy by absorbing more photons.  How these processes would offset each other and other factors governing star formation and the growth of the BH will be explored in future studies.

X-ray and SF feedback limit accretion to 0.2 - 0.8 $\dot{m}_{\mathrm{Edd}}$ in our models, rates at which quasars are not observed to have jets, so we do not include them in our models.  Steady jets are observed in active galactic nuclei (AGNe) at $L \lesssim 0.01 \, L_{\mathrm{Edd}}$ and intermittent jets are seen in quasars at $L \sim L_{\mathrm{Edd}}$ \citep{mh08}.  In these cases the disk is geometrically thick and radiatively inefficient.  No jets are seen at 0.01 $L_{\mathrm{Edd}} < L < L_{\mathrm{Edd}}$, when the disk is geometrically thin and radiatively efficient. 

While our numerical simulations show that J1120 can be explained as a DCBH that formed at the intersection of cold accretion streams at $z \sim$ 20, they also reveal that J1342 could have formed in the same way.  The BH reaches $1.18 \times 10^{8}$ \Ms\ by $z = 7.5$,  slightly above the error bar in observed J1342 mass but well within the uncertainty of a factor of three associated with the use of the Mg II line as a proxy for its luminosity. As shown in the right panel of Fig.~\ref{fig:bhmass}, we also obtain accretion rates of 27 \Ms\ yr$^{-1}$ at this redshift, in good agreement with observations.

Other scenarios could produce less massive BH seeds at high redshift.  For example, if a 10$^8$ \Ms\ halo is marginally enriched by a SN the gas can fragment locally, forming a dense nuclear cluster of stars that build up a single massive star via runaway collisions \citep{dv09,rein18,boek18}.  It then collapses to a BH that reaches 10$^3$ - 10$^4$ \Ms\ before the first SNe in the cluster begin to disrupt it \citep{wet08a}.  Some Pop III stars in less massive halos may reach masses up to 10$^3$ \Ms\ \citep{hir15}.  These processes could produce low-luminosity quasars at high redshift whose role in early reionization could be important, and they will be examined in future studies.  

\acknowledgments

We thank Fabio Pacucci, Britton Smith, Marta Volonteri, John Wise and Hao Xu for valuable discussions.  J. S., J. L. J. and H. L. were supported by a LANL LDRD Exploratory Research Grant 20170317ER.  D. J. W. was supported by STFC New Applicant Grant ST/P000509/1 and the European Research Council under the European Community's Seventh Framework Programme (FP7/2007 - 2013) via the ERC Advanced Grant "STARLIGHT:  Formation of the First Stars" (project number 339177).  Work at LANL was done under the auspices of the National Nuclear Security Administration of the U.S. Department of Energy at Los Alamos National Laboratory under Contract No. DE-AC52-06NA25396.  All Enzo models were performed under Institutional Computing (IC) allocations on Turquoise network platforms at LANL (Wolf).

\bibliographystyle{apj}
\bibliography{refs}

\begin{thebibliography}{96}
\expandafter\ifx\csname natexlab\endcsname\relax\def\natexlab#1{#1}\fi

\bibitem[{{Agarwal} {et~al.}(2012){Agarwal}, {Khochfar}, {Johnson}, {Neistein},
  {Dalla Vecchia}, \& {Livio}}]{agarw12}
{Agarwal}, B., {Khochfar}, S., {Johnson}, J.~L., {Neistein}, E., {Dalla
  Vecchia}, C., \& {Livio}, M. 2012, \mnras, 425, 2854

\bibitem[{{Alvarez} {et~al.}(2009){Alvarez}, {Wise}, \& {Abel}}]{awa09}
{Alvarez}, M.~A., {Wise}, J.~H., \& {Abel}, T. 2009, \apjl, 701, L133

\bibitem[{{Aykutalp} {et~al.}(2014){Aykutalp}, {Wise}, {Spaans}, \&
  {Meijerink}}]{aycin14}
{Aykutalp}, A., {Wise}, J.~H., {Spaans}, M., \& {Meijerink}, R. 2014, \apj,
  797, 139

\bibitem[{{Ba{\~n}ados} {et~al.}(2018){Ba{\~n}ados}, {Venemans},
  {Mazzucchelli}, {Farina}, {Walter}, {Wang}, {Decarli}, {Stern}, {Fan},
  {Davies}, {Hennawi}, {Simcoe}, {Turner}, {Rix}, {Yang}, {Kelson}, {Rudie}, \&
  {Winters}}]{ban18}
{Ba{\~n}ados}, E., {et~al.} 2018, \nat, 553, 473

\bibitem[{{Behroozi} \& {Silk}(2015)}]{bh14}
{Behroozi}, P.~S., \& {Silk}, J. 2015, \apj, 799, 32

\bibitem[{{Bieri} {et~al.}(2017){Bieri}, {Dubois}, {Rosdahl}, {Wagner}, {Silk},
  \& {Mamon}}]{bieri17}
{Bieri}, R., {Dubois}, Y., {Rosdahl}, J., {Wagner}, A., {Silk}, J., \& {Mamon},
  G.~A. 2017, \mnras, 464, 1854

\bibitem[{{Boekholt} {et~al.}(2018){Boekholt}, {Schleicher}, {Fellhauer},
  {Klessen}, {Reinoso}, {Stutz}, \& {Haemmerl{\'e}}}]{boek18}
{Boekholt}, T.~C.~N., {Schleicher}, D.~R.~G., {Fellhauer}, M., {Klessen},
  R.~S., {Reinoso}, B., {Stutz}, A.~M., \& {Haemmerl{\'e}}, L. 2018, \mnras,
  476, 366

\bibitem[{{Bournaud} {et~al.}(2011){Bournaud}, {Dekel}, {Teyssier}, {Cacciato},
  {Daddi}, {Juneau}, \& {Shankar}}]{bour11}
{Bournaud}, F., {Dekel}, A., {Teyssier}, R., {Cacciato}, M., {Daddi}, E.,
  {Juneau}, S., \& {Shankar}, F. 2011, \apjl, 741, L33

\bibitem[{{Bryan} {et~al.}(2014){Bryan}, {Norman}, {O'Shea}, {Abel}, {Wise},
  {Turk}, {Reynolds}, {Collins}, {Wang}, {Skillman}, {Smith}, {Harkness},
  {Bordner}, {Kim}, {Kuhlen}, {Xu}, {Goldbaum}, {Hummels}, {Kritsuk}, {Tasker},
  {Skory}, {Simpson}, {Hahn}, {Oishi}, {So}, {Zhao}, {Cen}, {Li}, \& {Enzo
  Collaboration}}]{enzo}
{Bryan}, G.~L., {et~al.} 2014, \apjs, 211, 19

\bibitem[{{Cen} \& {Ostriker}(1992)}]{co92}
{Cen}, R., \& {Ostriker}, J.~P. 1992, \apjl, 399, L113

\bibitem[{{Chen} {et~al.}(2017{\natexlab{a}}){Chen}, {Heger}, {Whalen},
  {Moriya}, {Bromm}, \& {Woosley}}]{chen17a}
{Chen}, K.-J., {Heger}, A., {Whalen}, D.~J., {Moriya}, T.~J., {Bromm}, V., \&
  {Woosley}, S.~E. 2017{\natexlab{a}}, \mnras, 467, 4731

\bibitem[{{Chen} {et~al.}(2017{\natexlab{b}}){Chen}, {Whalen}, {Wollenberg},
  {Glover}, \& {Klessen}}]{chen17b}
{Chen}, K.-J., {Whalen}, D.~J., {Wollenberg}, K.~M.~J., {Glover}, S.~C.~O., \&
  {Klessen}, R.~S. 2017{\natexlab{b}}, \apj, 844, 111

\bibitem[{{Chon} \& {Latif}(2017)}]{chon17a}
{Chon}, S., \& {Latif}, M.~A. 2017, \mnras, 467, 4293

\bibitem[{{Ciotti} {et~al.}(2009){Ciotti}, {Ostriker}, \& {Proga}}]{ciotti09}
{Ciotti}, L., {Ostriker}, J.~P., \& {Proga}, D. 2009, \apj, 699, 89

\bibitem[{{Costa} {et~al.}(2018{\natexlab{a}}){Costa}, {Rosdahl}, {Sijacki}, \&
  {Haehnelt}}]{costa18a}
{Costa}, T., {Rosdahl}, J., {Sijacki}, D., \& {Haehnelt}, M.~G.
  2018{\natexlab{a}}, \mnras, 473, 4197

\bibitem[{{Costa} {et~al.}(2018{\natexlab{b}}){Costa}, {Rosdahl}, {Sijacki}, \&
  {Haehnelt}}]{costa18b}
---. 2018{\natexlab{b}}, \mnras

\bibitem[{{Costa} {et~al.}(2014){Costa}, {Sijacki}, {Trenti}, \&
  {Haehnelt}}]{costa14}
{Costa}, T., {Sijacki}, D., {Trenti}, M., \& {Haehnelt}, M.~G. 2014, \mnras,
  439, 2146

\bibitem[{{Dalla Vecchia} \& {Schaye}(2012)}]{dvs12}
{Dalla Vecchia}, C., \& {Schaye}, J. 2012, \mnras, 426, 140

\bibitem[{{DeBuhr} {et~al.}(2010){DeBuhr}, {Quataert}, {Ma}, \&
  {Hopkins}}]{alphad}
{DeBuhr}, J., {Quataert}, E., {Ma}, C.-P., \& {Hopkins}, P. 2010, \mnras, 406,
  L55

\bibitem[{{Dekel} \& {Birnboim}(2006)}]{dek06}
{Dekel}, A., \& {Birnboim}, Y. 2006, \mnras, 368, 2

\bibitem[{{Dekel} {et~al.}(2009){Dekel}, {Birnboim}, {Engel}, {Freundlich},
  {Goerdt}, {Mumcuoglu}, {Neistein}, {Pichon}, {Teyssier}, \& {Zinger}}]{dek09}
{Dekel}, A., {et~al.} 2009, \nat, 457, 451

\bibitem[{{Devecchi} \& {Volonteri}(2009)}]{dv09}
{Devecchi}, B., \& {Volonteri}, M. 2009, \apj, 694, 302

\bibitem[{{Di Matteo} {et~al.}(2012){Di Matteo}, {Khandai}, {DeGraf}, {Feng},
  {Croft}, {Lopez}, \& {Springel}}]{dm12}
{Di Matteo}, T., {Khandai}, N., {DeGraf}, C., {Feng}, Y., {Croft}, R.~A.~C.,
  {Lopez}, J., \& {Springel}, V. 2012, \apjl, 745, L29

\bibitem[{{Dijkstra} {et~al.}(2014){Dijkstra}, {Ferrara}, \&
  {Mesinger}}]{dfm14}
{Dijkstra}, M., {Ferrara}, A., \& {Mesinger}, A. 2014, \mnras, 442, 2036

\bibitem[{{Done} {et~al.}(2012){Done}, {Davis}, {Jin}, {Blaes}, \&
  {Ward}}]{done12}
{Done}, C., {Davis}, S.~W., {Jin}, C., {Blaes}, O., \& {Ward}, M. 2012, \mnras,
  420, 1848

\bibitem[{{Dubois} {et~al.}(2015){Dubois}, {Volonteri}, {Silk}, {Devriendt},
  {Slyz}, \& {Teyssier}}]{dub15}
{Dubois}, Y., {Volonteri}, M., {Silk}, J., {Devriendt}, J., {Slyz}, A., \&
  {Teyssier}, R. 2015, \mnras, 452, 1502

\bibitem[{{Dunlop}(2013)}]{dunlop13}
{Dunlop}, J.~S. 2013, in Astrophysics and Space Science Library, Vol. 396, The
  First Galaxies, ed. T.~{Wiklind}, B.~{Mobasher}, \& V.~{Bromm}, 223

\bibitem[{{Feng} {et~al.}(2014){Feng}, {Di Matteo}, {Croft}, \&
  {Khandai}}]{yu14}
{Feng}, Y., {Di Matteo}, T., {Croft}, R., \& {Khandai}, N. 2014, \mnras, 440,
  1865

\bibitem[{{Ferland} {et~al.}(2013){Ferland}, {Porter}, {van Hoof}, {Williams},
  {Abel}, {Lykins}, {Shaw}, {Henney}, \& {Stancil}}]{cloudy}
{Ferland}, G.~J., {et~al.} 2013, {\em Rev. Mex. Astron \& Astrophys}, 49, 137

\bibitem[{{Glover} \& {Abel}(2008)}]{ga08}
{Glover}, S.~C.~O., \& {Abel}, T. 2008, \mnras, 388, 1627

\bibitem[{{Glover} \& {Jappsen}(2007)}]{japp07}
{Glover}, S.~C.~O., \& {Jappsen}, A.-K. 2007, \apj, 666, 1

\bibitem[{{Greif} {et~al.}(2011){Greif}, {White}, {Klessen}, \&
  {Springel}}]{greif11}
{Greif}, T.~H., {White}, S.~D.~M., {Klessen}, R.~S., \& {Springel}, V. 2011,
  \apj, 736, 147

\bibitem[{{Habouzit} {et~al.}(2017){Habouzit}, {Volonteri}, \&
  {Dubois}}]{hab17}
{Habouzit}, M., {Volonteri}, M., \& {Dubois}, Y. 2017, \mnras, 468, 3935

\bibitem[{{Haemmerl{\'e}} {et~al.}(2018{\natexlab{a}}){Haemmerl{\'e}}, {Woods},
  {Klessen}, {Heger}, \& {Whalen}}]{hle18}
{Haemmerl{\'e}}, L., {Woods}, T.~E., {Klessen}, R.~S., {Heger}, A., \&
  {Whalen}, D.~J. 2018{\natexlab{a}}, \apjl, 853, L3

\bibitem[{{Haemmerl{\'e}} {et~al.}(2018{\natexlab{b}}){Haemmerl{\'e}}, {Woods},
  {Klessen}, {Heger}, \& {Whalen}}]{hle17}
---. 2018{\natexlab{b}}, \mnras, 474, 2757

\bibitem[{{Hahn} \& {Abel}(2011)}]{hahn11}
{Hahn}, O., \& {Abel}, T. 2011, \mnras, 415, 2101

\bibitem[{{Hartwig} {et~al.}(2016){Hartwig}, {Latif}, {Magg}, {Bromm},
  {Klessen}, {Glover}, {Whalen}, {Pellegrini}, \& {Volonteri}}]{til15b}
{Hartwig}, T., {et~al.} 2016, \mnras, 462, 2184

\bibitem[{{Hirano} {et~al.}(2017){Hirano}, {Hosokawa}, {Yoshida}, \&
  {Kuiper}}]{hir17}
{Hirano}, S., {Hosokawa}, T., {Yoshida}, N., \& {Kuiper}, R. 2017, Science,
  357, 1375

\bibitem[{{Hirano} {et~al.}(2015){Hirano}, {Hosokawa}, {Yoshida}, {Omukai}, \&
  {Yorke}}]{hir15}
{Hirano}, S., {Hosokawa}, T., {Yoshida}, N., {Omukai}, K., \& {Yorke}, H.~W.
  2015, \mnras, 448, 568

\bibitem[{{Hirschmann} {et~al.}(2014){Hirschmann}, {Dolag}, {Saro}, {Bachmann},
  {Borgani}, \& {Burkert}}]{hir14}
{Hirschmann}, M., {Dolag}, K., {Saro}, A., {Bachmann}, L., {Borgani}, S., \&
  {Burkert}, A. 2014, \mnras, 442, 2304

\bibitem[{{Hosokawa} {et~al.}(2013){Hosokawa}, {Yorke}, {Inayoshi}, {Omukai},
  \& {Yoshida}}]{hos13}
{Hosokawa}, T., {Yorke}, H.~W., {Inayoshi}, K., {Omukai}, K., \& {Yoshida}, N.
  2013, \apj, 778, 178

\bibitem[{{Hummel} {et~al.}(2016){Hummel}, {Stacy}, \& {Bromm}}]{hum16}
{Hummel}, J.~A., {Stacy}, A., \& {Bromm}, V. 2016, \mnras, 460, 2432

\bibitem[{{Inayoshi} \& {Omukai}(2012)}]{io12}
{Inayoshi}, K., \& {Omukai}, K. 2012, \mnras, 422, 2539

\bibitem[{{Inayoshi} {et~al.}(2015){Inayoshi}, {Visbal}, \&
  {Kashiyama}}]{ivk15}
{Inayoshi}, K., {Visbal}, E., \& {Kashiyama}, K. 2015, \mnras, 453, 1692

\bibitem[{{Johnson} {et~al.}(2014){Johnson}, {Whalen}, {Agarwal},
  {Paardekooper}, \& {Khochfar}}]{jet14}
{Johnson}, J.~L., {Whalen}, D.~J., {Agarwal}, B., {Paardekooper}, J.-P., \&
  {Khochfar}, S. 2014, \mnras, 445, 686

\bibitem[{{Johnson} {et~al.}(2013){Johnson}, {Whalen}, {Li}, \& {Holz}}]{jet13}
{Johnson}, J.~L., {Whalen}, D.~J., {Li}, H., \& {Holz}, D.~E. 2013, \apj, 771,
  116

\bibitem[{{Kitayama} \& {Yoshida}(2005)}]{ky05}
{Kitayama}, T., \& {Yoshida}, N. 2005, \apj, 630, 675

\bibitem[{{Kuhlen} \& {Madau}(2005)}]{km05}
{Kuhlen}, M., \& {Madau}, P. 2005, \mnras, 363, 1069

\bibitem[{{Latif} {et~al.}(2014){Latif}, {Niemeyer}, \& {Schleicher}}]{lns14}
{Latif}, M.~A., {Niemeyer}, J.~C., \& {Schleicher}, D.~R.~G. 2014, \mnras, 440,
  2969

\bibitem[{{Latif} {et~al.}(2013){Latif}, {Schleicher}, {Schmidt}, \&
  {Niemeyer}}]{latif13a}
{Latif}, M.~A., {Schleicher}, D.~R.~G., {Schmidt}, W., \& {Niemeyer}, J. 2013,
  \mnras, 430, 588

\bibitem[{{Li} {et~al.}(2007){Li}, {Hernquist}, {Robertson}, {Cox}, {Hopkins},
  {Springel}, {Gao}, {Di Matteo}, {Zentner}, {Jenkins}, \& {Yoshida}}]{li07}
{Li}, Y., {et~al.} 2007, \apj, 665, 187

\bibitem[{{Lodato} \& {Natarajan}(2006)}]{ln06}
{Lodato}, G., \& {Natarajan}, P. 2006, \mnras, 371, 1813

\bibitem[{{Machacek} {et~al.}(2003){Machacek}, {Bryan}, \& {Abel}}]{mba03}
{Machacek}, M.~E., {Bryan}, G.~L., \& {Abel}, T. 2003, \mnras, 338, 273

\bibitem[{{Madau} {et~al.}(2014){Madau}, {Haardt}, \& {Dotti}}]{mhd14}
{Madau}, P., {Haardt}, F., \& {Dotti}, M. 2014, \apjl, 784, L38

\bibitem[{{Merloni} \& {Heinz}(2008)}]{mh08}
{Merloni}, A., \& {Heinz}, S. 2008, \mnras, 388, 1011

\bibitem[{{Milosavljevi{\'c}} {et~al.}(2009{\natexlab{a}}){Milosavljevi{\'c}},
  {Bromm}, {Couch}, \& {Oh}}]{milos09a}
{Milosavljevi{\'c}}, M., {Bromm}, V., {Couch}, S.~M., \& {Oh}, S.~P.
  2009{\natexlab{a}}, \apj, 698, 766

\bibitem[{{Milosavljevi{\'c}} {et~al.}(2009{\natexlab{b}}){Milosavljevi{\'c}},
  {Couch}, \& {Bromm}}]{milos09b}
{Milosavljevi{\'c}}, M., {Couch}, S.~M., \& {Bromm}, V. 2009{\natexlab{b}},
  \apjl, 696, L146

\bibitem[{{Mortlock} {et~al.}(2011){Mortlock}, {Warren}, {Venemans}, {Patel},
  {Hewett}, {McMahon}, {Simpson}, {Theuns}, {Gonz{\'a}les-Solares}, {Adamson},
  {Dye}, {Hambly}, {Hirst}, {Irwin}, {Kuiper}, {Lawrence}, \&
  {R{\"o}ttgering}}]{mort11}
{Mortlock}, D.~J., {et~al.} 2011, \nat, 474, 616

\bibitem[{{Nanni} {et~al.}(2017){Nanni}, {Vignali}, {Gilli}, {Moretti}, \&
  {Brandt}}]{nan17}
{Nanni}, R., {Vignali}, C., {Gilli}, R., {Moretti}, A., \& {Brandt}, W.~N.
  2017, \aap, 603, A128

\bibitem[{{O'Shea} {et~al.}(2005){O'Shea}, {Abel}, {Whalen}, \&
  {Norman}}]{oet05}
{O'Shea}, B.~W., {Abel}, T., {Whalen}, D., \& {Norman}, M.~L. 2005, \apjl, 628,
  L5

\bibitem[{{Pacucci} {et~al.}(2015){Pacucci}, {Ferrara}, {Volonteri}, \&
  {Dubus}}]{pac15}
{Pacucci}, F., {Ferrara}, A., {Volonteri}, M., \& {Dubus}, G. 2015, \mnras,
  454, 3771

\bibitem[{{Park} \& {Ricotti}(2011)}]{pm11}
{Park}, K., \& {Ricotti}, M. 2011, \apj, 739, 2

\bibitem[{{Park} \& {Ricotti}(2012)}]{pm12}
---. 2012, \apj, 747, 9

\bibitem[{{Park} \& {Ricotti}(2013)}]{pm13}
---. 2013, \apj, 767, 163

\bibitem[{{Planck Collaboration} {et~al.}(2016){Planck Collaboration}, {Ade},
  {Aghanim}, {Arnaud}, {Ashdown}, {Aumont}, {Baccigalupi}, {Banday},
  {Barreiro}, {Bartlett}, \& et~al.}]{planck2}
{Planck Collaboration} {et~al.} 2016, \aap, 594, A13

\bibitem[{{Reed} {et~al.}(2017){Reed}, {McMahon}, {Martini}, {Banerji},
  {Auger}, {Hewett}, {Koposov}, {Gibbons}, {Gonzalez-Solares}, {Ostrovski},
  {Tie}, {Abdalla}, {Allam}, {Benoit-L{\'e}vy}, {Bertin}, {Brooks},
  {Buckley-Geer}, {Burke}, {Carnero Rosell}, {Carrasco Kind}, {Carretero}, {da
  Costa}, {DePoy}, {Desai}, {Diehl}, {Doel}, {Evrard}, {Finley}, {Flaugher},
  {Fosalba}, {Frieman}, {Garc{\'{\i}}a-Bellido}, {Gaztanaga}, {Goldstein},
  {Gruen}, {Gruendl}, {Gutierrez}, {James}, {Kuehn}, {Kuropatkin}, {Lahav},
  {Lima}, {Maia}, {Marshall}, {Melchior}, {Miller}, {Miquel}, {Nord}, {Ogando},
  {Plazas}, {Romer}, {Sanchez}, {Scarpine}, {Schubnell}, {Sevilla-Noarbe},
  {Smith}, {Sobreira}, {Suchyta}, {Swanson}, {Tarle}, {Tucker}, {Walker}, \&
  {Wester}}]{reed17}
{Reed}, S.~L., {et~al.} 2017, \mnras, 468, 4702

\bibitem[{{Regan} \& {Downes}(2018)}]{rd18}
{Regan}, J.~A., \& {Downes}, T.~P. 2018, \mnras, 475, 4636

\bibitem[{{Regan} \& {Haehnelt}(2009)}]{rh09b}
{Regan}, J.~A., \& {Haehnelt}, M.~G. 2009, \mnras, 396, 343

\bibitem[{{Regan} {et~al.}(2017){Regan}, {Visbal}, {Wise}, {Haiman},
  {Johansson}, \& {Bryan}}]{regan17a}
{Regan}, J.~A., {Visbal}, E., {Wise}, J.~H., {Haiman}, Z., {Johansson}, P.~H.,
  \& {Bryan}, G.~L. 2017, Nature Astronomy, 1, 0075

\bibitem[{{Reinoso} {et~al.}(2018){Reinoso}, {Schleicher}, {Fellhauer},
  {Klessen}, \& {Boekholt}}]{rein18}
{Reinoso}, B., {Schleicher}, D.~R.~G., {Fellhauer}, M., {Klessen}, R.~S., \&
  {Boekholt}, T.~C.~N. 2018, \aap, 614, A14

\bibitem[{{Ritter} {et~al.}(2012){Ritter}, {Safranek-Shrader}, {Gnat},
  {Milosavljevi{\'c}}, \& {Bromm}}]{ritt12}
{Ritter}, J.~S., {Safranek-Shrader}, C., {Gnat}, O., {Milosavljevi{\'c}}, M.,
  \& {Bromm}, V. 2012, \apj, 761, 56

\bibitem[{{Ritter} {et~al.}(2016){Ritter}, {Safranek-Shrader},
  {Milosavljevi{\'c}}, \& {Bromm}}]{rit16}
{Ritter}, J.~S., {Safranek-Shrader}, C., {Milosavljevi{\'c}}, M., \& {Bromm},
  V. 2016, \mnras, 463, 3354

\bibitem[{{Safranek-Shrader} {et~al.}(2014){Safranek-Shrader},
  {Milosavljevi{\'c}}, \& {Bromm}}]{ss13}
{Safranek-Shrader}, C., {Milosavljevi{\'c}}, M., \& {Bromm}, V. 2014, \mnras,
  438, 1669

\bibitem[{{Schaerer}(2002)}]{s02}
{Schaerer}, D. 2002, \aap, 382, 28

\bibitem[{{Schauer} {et~al.}(2017){Schauer}, {Regan}, {Glover}, \&
  {Klessen}}]{srg17}
{Schauer}, A.~T.~P., {Regan}, J., {Glover}, S.~C.~O., \& {Klessen}, R.~S. 2017,
  \mnras, 471, 4878

\bibitem[{{Shang} {et~al.}(2010){Shang}, {Bryan}, \& {Haiman}}]{sbh10}
{Shang}, C., {Bryan}, G.~L., \& {Haiman}, Z. 2010, \mnras, 402, 1249

\bibitem[{{Sluder} {et~al.}(2016){Sluder}, {Ritter}, {Safranek-Shrader},
  {Milosavljevi{\'c}}, \& {Bromm}}]{slud16}
{Sluder}, A., {Ritter}, J.~S., {Safranek-Shrader}, C., {Milosavljevi{\'c}}, M.,
  \& {Bromm}, V. 2016, \mnras, 456, 1410

\bibitem[{{Smidt} {et~al.}(2016){Smidt}, {Wiggins}, \& {Johnson}}]{smidt16a}
{Smidt}, J., {Wiggins}, B.~K., \& {Johnson}, J.~L. 2016, \apjl, 829, L6

\bibitem[{{Smith} {et~al.}(2018){Smith}, {Regan}, {Downes}, {Norman}, {O'Shea},
  \& {Wise}}]{srd18}
{Smith}, B., {Regan}, J., {Downes}, T., {Norman}, M., {O'Shea}, B., \& {Wise},
  J. 2018, arXiv:1804.06477

\bibitem[{{Smith} {et~al.}(2015){Smith}, {Wise}, {O'Shea}, {Norman}, \&
  {Khochfar}}]{brit15}
{Smith}, B.~D., {Wise}, J.~H., {O'Shea}, B.~W., {Norman}, M.~L., \& {Khochfar},
  S. 2015, \mnras, 452, 2822

\bibitem[{{Sobral} {et~al.}(2015){Sobral}, {Matthee}, {Darvish}, {Schaerer},
  {Mobasher}, {R{\"o}ttgering}, {Santos}, \& {Hemmati}}]{cr7}
{Sobral}, D., {Matthee}, J., {Darvish}, B., {Schaerer}, D., {Mobasher}, B.,
  {R{\"o}ttgering}, H.~J.~A., {Santos}, S., \& {Hemmati}, S. 2015, \apj, 808,
  139

\bibitem[{{Stacy} {et~al.}(2011){Stacy}, {Bromm}, \& {Loeb}}]{stacy11a}
{Stacy}, A., {Bromm}, V., \& {Loeb}, A. 2011, \apjl, 730, L1

\bibitem[{{Trenti} {et~al.}(2008){Trenti}, {Santos}, \& {Stiavelli}}]{tss08}
{Trenti}, M., {Santos}, M.~R., \& {Stiavelli}, M. 2008, \apj, 687, 1

\bibitem[{{Umeda} {et~al.}(2016){Umeda}, {Hosokawa}, {Omukai}, \&
  {Yoshida}}]{um16}
{Umeda}, H., {Hosokawa}, T., {Omukai}, K., \& {Yoshida}, N. 2016, \apjl, 830,
  L34

\bibitem[{{Venemans} {et~al.}(2017){Venemans}, {Walter}, {Decarli},
  {Ba{\~n}ados}, {Hodge}, {Hewett}, {McMahon}, {Mortlock}, \&
  {Simpson}}]{vet17}
{Venemans}, B.~P., {et~al.} 2017, \apj, 837, 146

\bibitem[{{Volonteri} {et~al.}(2015){Volonteri}, {Silk}, \& {Dubus}}]{vsd15}
{Volonteri}, M., {Silk}, J., \& {Dubus}, G. 2015, \apj, 804, 148

\bibitem[{{Whalen} {et~al.}(2004){Whalen}, {Abel}, \& {Norman}}]{wan04}
{Whalen}, D., {Abel}, T., \& {Norman}, M.~L. 2004, \apj, 610, 14

\bibitem[{{Whalen} {et~al.}(2010){Whalen}, {Hueckstaedt}, \&
  {McConkie}}]{wet10}
{Whalen}, D., {Hueckstaedt}, R.~M., \& {McConkie}, T.~O. 2010, \apj, 712, 101

\bibitem[{{Whalen} {et~al.}(2008{\natexlab{a}}){Whalen}, {O'Shea}, {Smidt}, \&
  {Norman}}]{wet08b}
{Whalen}, D., {O'Shea}, B.~W., {Smidt}, J., \& {Norman}, M.~L.
  2008{\natexlab{a}}, \apj, 679, 925

\bibitem[{{Whalen} {et~al.}(2008{\natexlab{b}}){Whalen}, {van Veelen},
  {O'Shea}, \& {Norman}}]{wet08a}
{Whalen}, D., {van Veelen}, B., {O'Shea}, B.~W., \& {Norman}, M.~L.
  2008{\natexlab{b}}, \apj, 682, 49

\bibitem[{{Whalen} \& {Fryer}(2012)}]{wf12}
{Whalen}, D.~J., \& {Fryer}, C.~L. 2012, \apjl, 756, L19

\bibitem[{{Wise} \& {Abel}(2011)}]{moray}
{Wise}, J.~H., \& {Abel}, T. 2011, \mnras, 414, 3458

\bibitem[{{Wise} {et~al.}(2008){Wise}, {Turk}, \& {Abel}}]{wta08}
{Wise}, J.~H., {Turk}, M.~J., \& {Abel}, T. 2008, \apj, 682, 745

\bibitem[{{Woods} {et~al.}(2017){Woods}, {Heger}, {Whalen}, {Haemmerl{\'e}}, \&
  {Klessen}}]{tyr17}
{Woods}, T.~E., {Heger}, A., {Whalen}, D.~J., {Haemmerl{\'e}}, L., \&
  {Klessen}, R.~S. 2017, \apjl, 842, L6

\bibitem[{{Xu} {et~al.}(2014){Xu}, {Ahn}, {Wise}, {Norman}, \& {O'Shea}}]{1kev}
{Xu}, H., {Ahn}, K., {Wise}, J.~H., {Norman}, M.~L., \& {O'Shea}, B.~W. 2014,
  \apj, 791, 110

\bibitem[{{Yue} {et~al.}(2014){Yue}, {Ferrara}, {Salvaterra}, {Xu}, \&
  {Chen}}]{yue14}
{Yue}, B., {Ferrara}, A., {Salvaterra}, R., {Xu}, Y., \& {Chen}, X. 2014,
  \mnras, 440, 1263

\end{thebibliography}

\end{document}